\documentclass[aps,preprint,superscriptaddress]{revtex4-2}

\usepackage{graphicx}

\begin{document}

\title{Quasispecies relocate rather than drift: hierarchical genotype networks track viral adaptation}

\author{Iker Atienza-Diez}
\affiliation{Museo Nacional de Ciencias Naturales (CSIC), Madrid, Spain}
\affiliation{Grupo Interdisciplinar de Sistemas Complejos (GISC), Madrid, Spain}

\author{Samuel Martínez-Alcalá}
\affiliation{Museo Nacional de Ciencias Naturales (CSIC), Madrid, Spain}
\affiliation{Grupo Interdisciplinar de Sistemas Complejos (GISC), Madrid, Spain}

\author{Luis F. Seoane}
\affiliation{Institut de Biologia Evolutiva, CSIC-UPF, Pg Maritim de la Barceloneta 37, 08003 Barcelona, Spain}
\affiliation{Theoretical Sciences Visiting Program, Okinawa Institute of Science and Technology Graduate University, Onna, 904-0495, Japan}

\author{Susanna Manrubia}
\email{smanrubia@mncn.csic.es}
\affiliation{Museo Nacional de Ciencias Naturales (CSIC), Madrid, Spain}
\affiliation{Grupo Interdisciplinar de Sistemas Complejos (GISC), Madrid, Spain}

\begin{abstract}
Viral populations are ensembles of mutationally related genotypes that explore sequence space locally. Their genotype networks are hierarchical: abundance decays with mutational distance from the most abundant genotype (the root) and degree distributions are heavy-tailed. How this architecture responds to selective challenges remains unclear. We combine simulations of sequence populations on NK fitness landscapes with longitudinal hepatitis B virus (HBV) deep-sequencing data from chronically infected patients under treatment. Quasispecies structure is rapidly rebuilt around the current root, while root replacement is occasional, when a fitter variant is found. Adaptation unfolds on two timescales: fast, local reorganisation and slow, stochastic relocation. HBV populations of patients not responding to therapy show stasis. Meanwhile, responders show either near-complete replacement by a new root, or transient excursions followed by return to the original root. These patterns resemble the stasis and relocation regimes found in simulations. Our results suggest that genotype network organization and dynamics under selective pressure present generic features of molecular adaptation, with a potentially informative clinical readout.
\end{abstract}

\maketitle

\section{Introduction}

Viral populations are heterogeneous ensembles of mutationally related genotypes, the natural counterpart of molecular quasispecies \cite{eigen:1971,domingo:2019}. High mutation rates and large population sizes allow such clouds to generate large amounts of diversity, as repeatedly observed in RNA viruses \cite{andino:2015}, and therefore to explore sequence space continuously. Deep-sequencing techniques now make it possible to resolve viral diversity at the level of individual genotype fragments of a few hundred nucleotides \cite{beerenwinkel:2011}. This has opened the possibility of exploring the structure and organization of viral quasispecies to an unprecedented level of detail: genotype networks can be reconstructed by representing each genotype as a node and linking genotypes that differ by a single mutation. The topology of such networks summarises how the population explores and occupies sequence space \cite{seoane:2026,martinezalcala:2026}.
 
Genotype spaces are astronomically large, yet viral populations sample only a minute neighbourhood of possible genotypes \cite{louis:2016,manrubia:2026}. Theories based on neutral networks and holey landscapes \cite{schuster:1994b,huynen:1996b,gavrilets:2004} have put forward possible mechanisms to explain how evolution could traverse such spaces, emphasising global connectivity and navigability \cite{srivastava:2026}. Recent work instead points to local search: mutation-driven exploration around abundant sequences generates large diversity and phenotypic innovation without requiring a global search. The high dimensionality of sequence space allows the discovery of new viable phenotypes often just a few mutations away from extant genotypes \cite{gruner:1996b}. In this way, abundant phenotypes bias evolutionary trajectories by selecting functional solutions that can be attained through massive exploration of nearby variants \cite{manrubia:2026}. Consistently, genotype networks reconstructed from Q$\beta$ phage \cite{seoane:2026} and SARS-CoV-2 \cite{martinezalcala:2026} populations display a hierarchical structure that results from replication and mutation around each genotype; this structure is dominated by a root, the momentarily most abundant sequence in the population. Starting at the root, abundance decays roughly exponentially with mutational distance, showing little signature of extended neutral wandering far beyond the root as the mechanism behind viral adaptation.

So far, studies have been mostly focused on describing the architecture of quasispecies. Viral dynamics, however, are much less explored. Natural populations face challenges such as changes in temperature, immune responses or antiviral treatment, which alter the fitness landscape. If search is local, adaptation to a challenge may require that the root moves and that the mutant cloud reorganizes. Based on numerical simulations \cite{krug:2005} and on our observations with reconstructed genotype networks \cite{seoane:2026}, we hypothesise that these processes operate on two different timescales: fast, local rebuilding of the hierarchical network around the fittest nearby genotype, and slow, stochastic relocation when a new local maximum is found. This picture refines the notion of punctuated change \cite{eldredge:1972} to a resolution finer than consensus-sequence turnover, and offers a molecular counterpart of shifting-balance ideas \cite{wright:1932,devisser:2014} by suggesting that fitter variants are located through the massive generation of diversity near highly abundant genotypes, thus traversing low-fitness variants without the need to use neutral, possibly longer, detours in sequence space.

Here we test this hypothesis in simulations and \emph{in vivo}. We first simulate populations on NK landscapes \cite{kauffman:1987} under different initial conditions, and identify the signatures of stasis and relocation of the quasispecies in sequence space by reconstructing genotype networks in such synthetic populations. We then analyse longitudinal deep-sequencing data from the major hydrophilic region of the hepatitis B virus (HBV) surface antigen in chronically infected patients treated with the nucleic acid polymer REP~2139-Ca \cite{almahtab:2016,usman:2019}. HBV is a DNA virus replicating through reverse transcription, so it differs from the RNA viruses previously analyzed at this scale \cite{seoane:2026,martinezalcala:2026}. The data set we use corresponds to a natural experiment with different clinical outcomes that can be partly interpreted in the framework of synthetic quasispecies dynamics. We ask whether changes in quasispecies location and structure accompany treatment outcome, and discuss the potential of reconstructed genotype networks as a clinically informative quantity.

\section{Materials and Methods}
\label{sec:MM}

\subsection{Numerical simulations}
\label{sec:MM-a}

Genotypes are sequences of symbols drawn from a four-letter alphabet, $g\in\{A, C, G, U\}^L$. Fitness is generated through an NK model \cite{kauffman:1987}: $F(g)=L^{-1}\sum_{i=1}^{L} f_i(g_i;g_{i_1},\dots,g_{i_K})$, where each site's contribution depends on its own state and that of $K$ other sites, and the $4^{K+1}$ values of each $f_i$ are drawn independently from $[0,1]$. Parameter $K$ tunes ruggedness, from a single smooth peak ($K=0$) to uncorrelated fitness ($K=L-1$). We use $L=250$ and $K=4$ as representative of generic rough, correlated landscapes, with no {\it a priori} attempt to quantitatively fit empirical observations. 

We also make our landscape {\em holey} by including a fraction of lethal mutations. We take a linear fitness threshold, $f_\ell$, so that the fitness of a sequence becomes $F(g) \leftarrow \max[0,F(g)-f_\ell]$. All sequences initially with $F(g)\le f_\ell$ are thus non-viable \cite{aguirre:2015,yubero:2017}. For each quenched landscape, $f_\ell$ is calibrated such that the least-fit starting condition ($g_{\rm l}$, see below) has a reproductive fitness $R_{\rm l}=e^{2\lambda}\simeq1.73$ times the average among its $3L$ one-mutant neighbors. $\lambda$ is the parameter of a Poisson distribution setting the mean number of mutations per copied sequence. These choices keep the population in a localized quasispecies with a sizeable mutational cloud---since for the exact copy probability ($p_0=e^{-\lambda}$) we get $p_0R_{\rm l}=e^{\lambda}>1$. This is not an error-threshold condition, but a modeling convention that places different realizations in a comparable regime. $f_\ell$ is kept fixed when different initial populations are compared on the same landscape.

A population of $N=10^6$ sequences evolves under Wright--Fisher replacement, with offspring sampled proportionally to genotype fitness; each copy carries a Poisson-distributed number of mutations with $\lambda=0.275$, at uniformly random sites. Runs start from a clonal population at a root genotype chosen in one of three ways: \emph{global}, $g_{\rm g}$, the exact global maximum; \emph{evolved}, $g_{\rm e}$, the fittest genotype found within $5\,000$ generations; and \emph{local}, $g_{\rm l}$, a local maximum (all $3L$ one-mutant neighbours have lower fitness) reached from a random starting point. We sampled our simulations for network visualization and analysis after $t=50$, $100$, and $200$ generations. Codes to replicate our results are available in \cite{SamMAPP_QuasispeciesCFL}. 

\subsection{HBV data}

\begin{figure}[t]
    \centering
    \includegraphics[width=\linewidth]{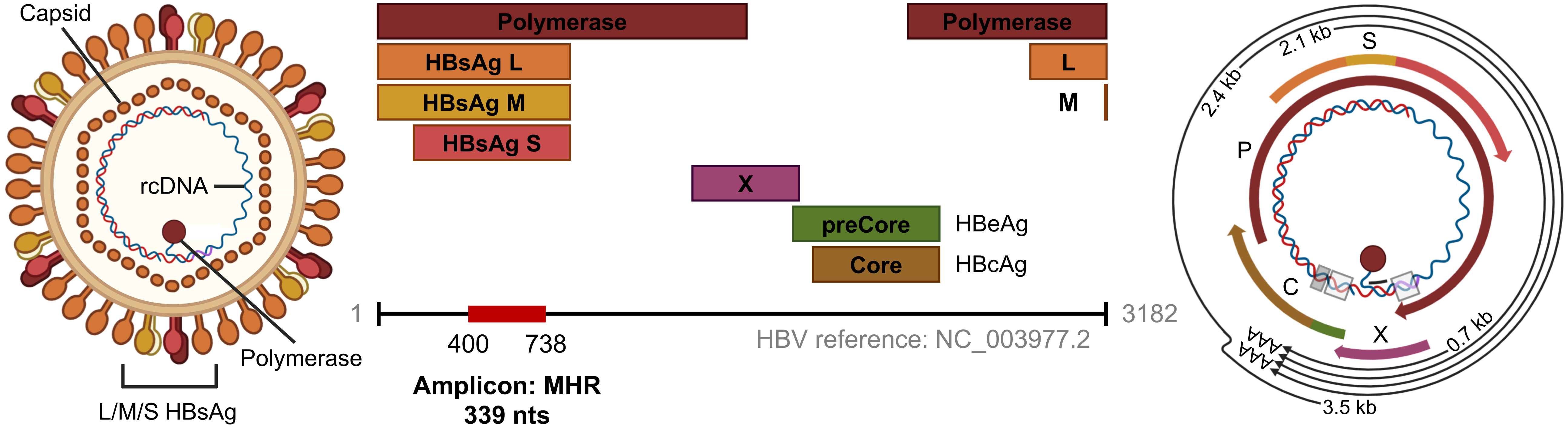}
    \caption{\textbf{Hepatitis B virus genome.} The hepatitis B virus (HBV) has a short, partially double-stranded, relaxed-circular DNA (rcDNA) genome of approximately 3200 nucleotides. The genome is composed of a full-length negative strand and a shorter, variable-length, positive strand that contains four overlapping open reading frames that encode seven viral proteins. The C gene encodes the internal nucleocapsid HBV core protein (HBcAg) along with its secreted pre-Core form (HBeAg). The S gene encodes the three viral envelope glycoproteins: large (L-), medium (M-) and small (S-) surface antigens (HBsAg). The P gene encodes the large viral DNA polymerase with reverse transcriptase activity that allows the virus to complete its complex cycle, generating the first strand of the DNA genome from an RNA intermediate. The X gene encodes the HBx regulatory protein, which modulates host transcription and plays a key role in viral replication. The HBV amplicon analyzed in this work is 339 nucleotides long and spans nucleotides 400 to 738, both included (positions relative to reference genome with NCBI accession number NC\_003977.2). Virion structure and circular genome organization diagrams were modified from BioRender \cite{bradley:2026}. 
    }
    \label{fig:HBV-genome}
\end{figure}

The hepatitis B virus (HBV) has a short DNA genome of approximately 3200 nucleotides (Figure~\ref{fig:HBV-genome}). In humans, HBV specifically infects hepatocytes and causes serious liver diseases, such as acute or chronic hepatitis and cirrhosis. We use previously published data of patients with chronic HBV infection \cite{almahtab:2016,usman:2019}, re-analysing deep-sequencing data from a subsample of eight (out of twelve) HBeAg-positive treatment-naive patients (labeled REP1--6, REP8, REP11) of the REP102 trial. All patients were under therapy with the nucleic acid polymer REP~2139-Ca \cite{almahtab:2016,usman:2019} (Table~\ref{tab:patient_summary}). Patient serum was sampled three times (S1--S3) at patient-specific weeks. Responders and non-responders were classified as in \cite{usman:2019}. The sequenced amplicon (339~nt, positions 400--738 of NC\_003977.2) covers the major hydrophilic region of the HBsAg protein, including the ``a'' determinant region---the main target of neutralising antibodies, thus under constant immune pressure (Fig. \ref{fig:HBV-genome}). 

\begin{table}[!h]
    \centering
    \caption{Clinical summary: Sampling is pre-treatment (PT) or weeks relative to REP~2139-Ca therapy starting date (e.g.\ W5). Information summarized from clinical trial REP102 \cite{almahtab:2016,usman:2019}. Non-responders (criteria in \cite{usman:2019}) are italicized.}
    \label{tab:patient_summary}
    \begin{tabular}{p{1.3cm}p{2.5cm}p{6.5cm}}
        \hline
        {\bf Patient} &
        {\bf Sampling times} &
        {\bf HBV DNA and HBsAg markers} \\
        
        \hline
        {\em REP1} & 
        {\em PT, W26, W39} &
        {\em Persistently high.} \\
        
        REP2 &
        PT, W5, W11 &
        Progressive reduction. \\
        
        REP3 &
        PT, W5, W16 &
        Marked reduction during early treatment. \\
        
        REP4 &
        PT, W12, W26 &
        Marked reduction during treatment. \\
        
        {\em REP5} &
        {\em W2, W23, W34\textsuperscript{a}} &
        {\em Limited reduction.} \\
        
        REP6 &
        PT, W13, W29 &
        Substantial declines at intermediate time. \\
        
        REP8 &
        W1, W10, W19 &
        Early suppression. \\
        
        REP11 &
        PT, W7, W12 &
        Rapid and pronounced reduction at early time. \\
        
        \hline
    \end{tabular}
    \begin{flushleft}
        \footnotesize
        \textsuperscript{a} Thymosin~$\alpha$1 added to treatment between second and third sampling points.
    \end{flushleft}
\end{table}

\subsection{Network reconstruction and observables}
\label{sec:sims_obs}
Raw sequencing data were downloaded from SRA NCBI accession number SRP154287 (BioProject PRJNA478149). Raw reads were merged, aligned to the reference and filtered for misaligned, indel-containing and low-quality sequences, and collapsed into genotypes with their abundances, as in \cite{seoane:2026,martinezalcala:2026,villanueva:2022}. Genotype networks link genotypes differing by one nucleotide. Only the largest connected component (LCC) is kept, representing a severe and consistent filter for mutationally coupled variants. Network reconstruction was performed using a pivot-based metric-similarity algorithm, publicly available along with instructions \cite{pivonet}. Simulated populations are processed identically. Since they are one order of magnitude larger than deep-sequenced sets, network visualizations from simulations only show nodes with abundance $5$ or higher. 

We took a series of measurements to characterize our networks: (i) {\em population overlaps} between different sampling times, represented as fractions in Venn diagrams and as a function of sequence abundance; (ii) {\em abundance profiles} measured as fractions of sequences found at a given distance from the initial root (i.e., from $g_{\rm g}$, $g_{\rm e}$, or $g_{\rm l}$ in simulations, or from the S1 root in HBV data); (iii) {\em cosine distance}, normalizing abundances as unitary vectors, $\mathbf{v}_{i}$, with $i$ indexing sampling times ($t=50$, $100$, and $200$ or S1--3):  
\begin{equation}
\cos\varphi_{ij}=\frac{\mathbf{v}_i\cdot\mathbf{v}_j}{\|\mathbf{v}_i\|\,\|\mathbf{v}_j\|},\qquad 0\le\varphi_{ij}\le\pi/2. 
\end{equation}
Abundance distributions (not shown) are heavy-tailed, hence $v_{i}$ and $\cos\varphi_{ij}$ are dominated by the persistence (or not) of the population root. Turnover of the dominant phenotype is usually enough to collapse $\cos\varphi_{ij}$ to $\simeq0$. Meanwhile, we only take sequence presence or absence to compute overlaps (either as Venn diagrams or otherwise). This allows us to compare population composition beyond the root---distinguishing, e.g., periphery renewal. Both approaches are necessary to understand heavy-tailed populations, in which neither the average nor the extreme of the distributions give a complete picture. 

\section{Simulations: a traveling localized quasispecies}
\label{sec:sims}

\begin{figure}[t]
    \centering
    \includegraphics[width=0.8\linewidth]{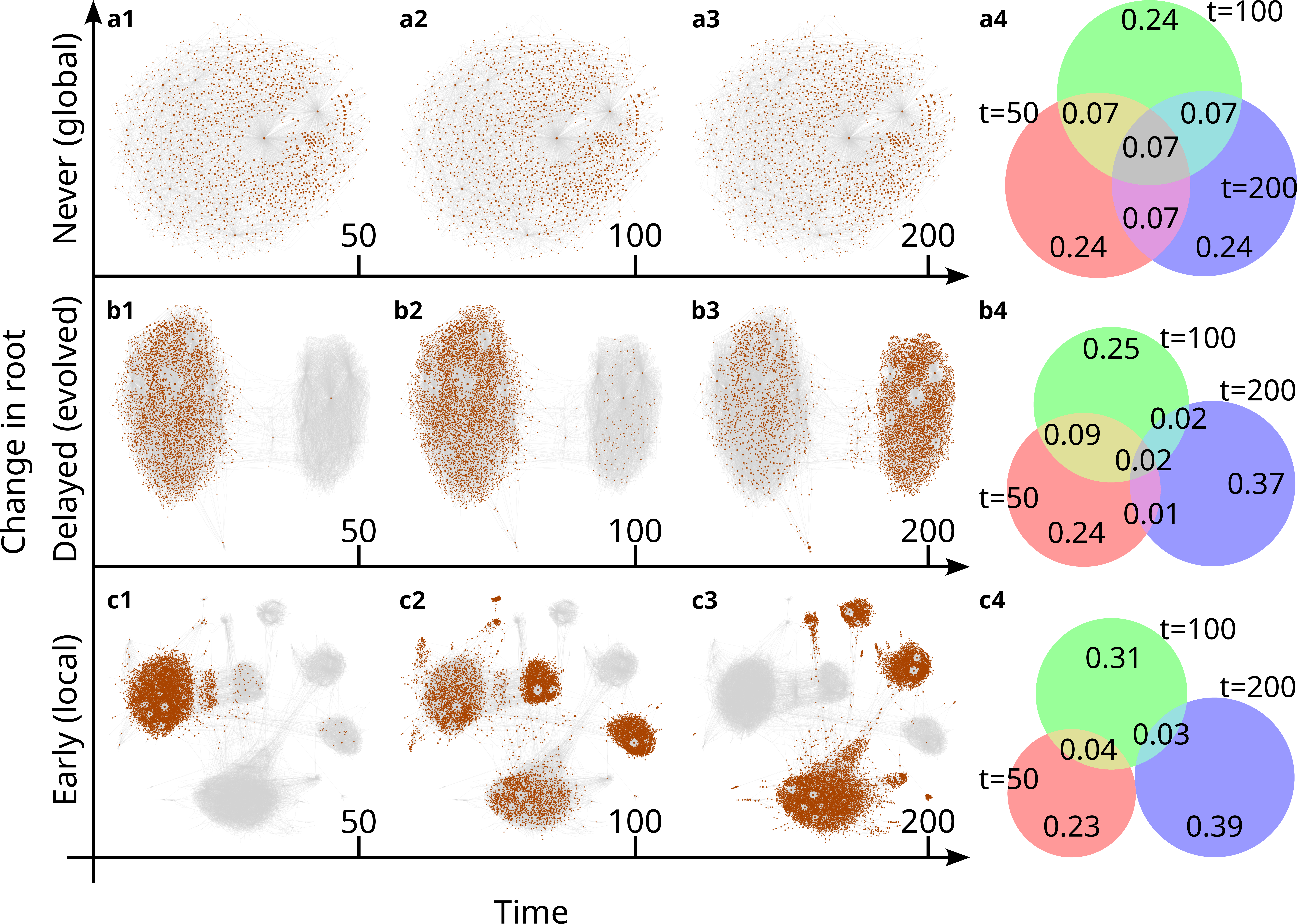}
    
    \caption{{\bf Genotype networks from numerical simulations.} We instantiated three simulations on a same fitness landscape starting from three different monoclonal populations---in order of decreasing initial adaptation: ({\bf a}) the global optimum, $g_{\rm g}$; ({\bf b}) an evolved fit sequence, $g_{\rm e}$; and ({\bf c}) a local maximum reached from a random sequence, $g_{\rm l}$. This ordering also induces an axis (vertical) of how rapidly the simulated quasispecies moves away from the initial condition (to be compared with how rapidly HBV patients respond to treatment, Fig.\ \ref{fig:HBV-nets}). Nodes with non-zero abundance are shown in orange at $t=50$ ({\bf a1}, {\bf b1}, and {\bf c1}), $t=100$ ({\bf a2}, {\bf b2}, and {\bf c2}), and $t=200$ ({\bf a3}, {\bf b3}, and {\bf c3}). Node size is proportional to the logarithm of the abundance in each case. All edges (even to inactive nodes) show the network layout at all times. Venn diagrams ({\bf a4}, {\bf b4}, and {\bf c4}) show the fraction of all sequences that are present at each individual sampling, and at intersections between samplings (note that the intersection between two samplings must include the fraction present in all three).}

    \label{fig:Num-nets}
\end{figure}

We use numerical simulations as a controlled setting, with a known, static fitness landscape, to identify stasis and relocation of an ensemble of replicators. The three instances in Fig.~\ref{fig:Num-nets} run over NK landscapes and differ in the initial root's fitness---i.e., in how well adapted the population starts (Section \ref{sec:MM-a}). Rather than an external perturbation (e.g.\ a therapy), the evolutionary ``challenge'' is the distance between that initial root and fitter genotypes available. We show networks after $t=50$, $100$, and $200$ generations. The initial clone develops a hierarchical cloud characteristic of quasispecies \cite{seoane:2026,martinezalcala:2026} within a few generations. From there, simulations differ depending on the starting condition: 

\begin{figure}[t]
    \centering
    
    \includegraphics[width=\linewidth]{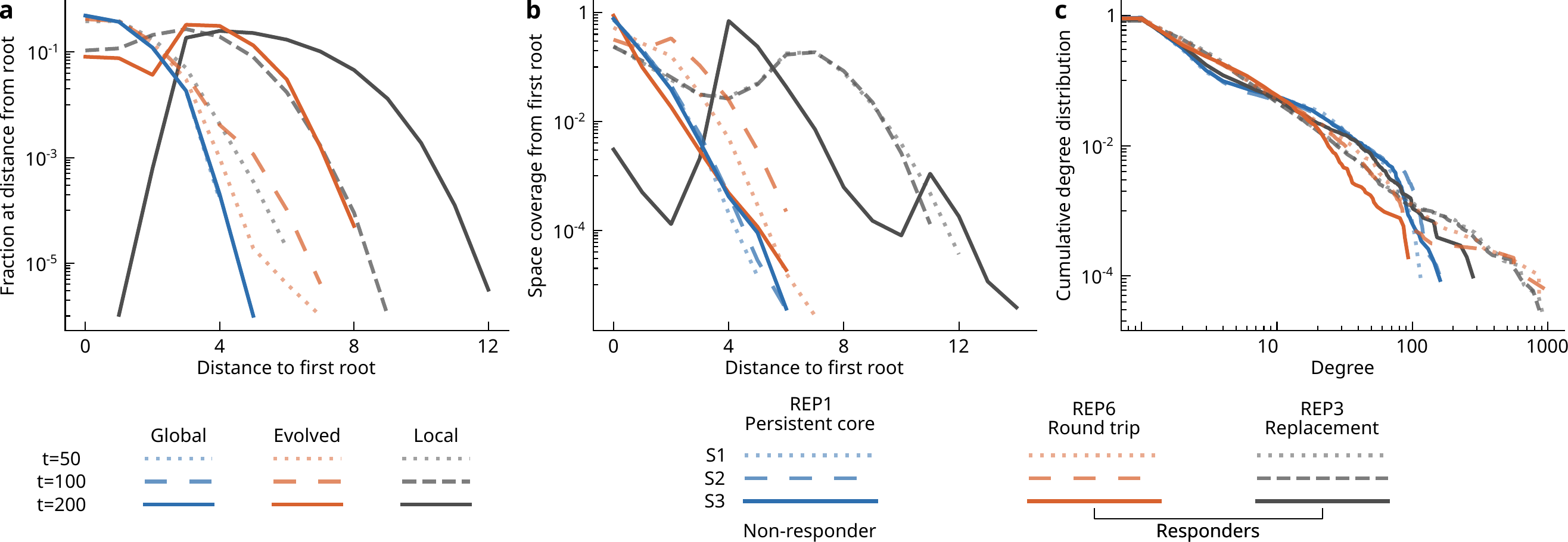}
    \caption{{\bf Displacement of the quasispecies relative to its initial root and degree distribution of empirical networks.} ({\bf a-b}) Abundance profiles (Sect.\ \ref{sec:sims_obs}) tell the fraction of sequences in a sample found at each distance from the root of the earliest sample. The shifting distributions tell a story that suggests possible dynamical classes. ({\bf a}) Abundance profiles in simulated networks stay in place for global optima as starting conditions (blue curves collapsed on top of each other), revealing a persistent cloud around the stable root. For evolved initial conditions (orange) the profiles persist in the first two samples (discontinuous) but depart at the last sample (solid) indicating a change of root. Local optima (gray) present more immediate relocation, and abundance profiles diverge immediately. ({\bf b}) Abundance profiles for empirical HBV show similarities with simulations (a stable peak for non-responder patient REP1, blue; profiles that escape the root for early responder patient REP3, gray), but also dissimilarities (a departure, dashed orange, followed by re-assembly closer to the root, solid orange, for delayed responder patient REP6). ({\bf c}) All empirical degree distributions display heavy tails and collapse rather neatly, indicating shared properties despite differences in structure, root, or response to treatment. }
    \label{fig:sim_distances}
\end{figure}

\emph{Global root.} When the population starts at the global optimum, any mutant is less fit, and the root remains the most abundant genotype throughout (Fig.~\ref{fig:Num-nets}a1-3). All three abundance profiles are similar (blue curves in Fig.\ \ref{fig:sim_distances}a), with abundance decaying fast with distance to the initial root, $g_{\rm g}$. Regardless, mutation continuously generates a large cloud of variants and only a small core ($\sim7\%$ of sequences) is shared over time (Fig.~\ref{fig:Num-nets}a4). The low-abundance periphery is stochastically renewed between samples. Consider the huge numbers of existing sequences, from $\sim7 \times 10^7$ at distance $d=3$ from the root to $10^{16}$ at $d=7$. A population sized $N=10^6$ necessarily produces previously unexplored diversity at each generation. Stasis of the root thus coexists with sustained diversity and a fluctuating periphery, a signature of replication-exploration balance around a fixed peak \cite{seoane:2026}. 

\emph{Evolved root.} Starting from the fittest genotype found within $5,000$ generations, $g_{\rm e}$, the initial fitness is high but not maximal. Rare, fitter variants eventually emerge at the periphery. For a long period, the dynamics resemble the global case (Fig.\ \ref{fig:Num-nets}{b1-2}): around $11\%$ of sequences are shared between $t=50$ and $100$ (Fig.\ \ref{fig:Num-nets}b4), the root persists, and the network keeps its hierarchical organisation around $g_{\rm e}$ (discontinuous orange curves in Fig.\ \ref{fig:sim_distances}a). But after long enough, the population relocates (Fig.\ \ref{fig:Num-nets}{b3}). Its hierarchical structure is readily rebuilt around a different root $3$ mutations away from $g_{\rm e}$ (solid orange curve in Fig.\ \ref{fig:sim_distances}a). The abundance profile was already broader at $t=100$ than at $t=50$ (dashed vs.\ dotted orange curves at Fig.\ \ref{fig:sim_distances}a), suggesting that the new quasispecies was already getting assembled at the periphery before taking over. Simulations started at $g_{\rm e}$ are compatible with a metastable state with relatively long stasis, eventually replaced by a fitter solution still far from the global maximum. The time needed to find that new root is much longer than the time to rebuild the structure around it.

\emph{Local root.} Starting at a local maximum reached from a random genotype, $g_{\rm l}$, fitness is low and many fitter peaks lie within quasispecies reach. The population changes composition rapidly as new local maxima are found. At $t=100$ the network shows several ``communities'' hierarchically structured around coexisting hubs (Fig.~\ref{fig:Num-nets}c2). At $t=200$, nearly 90\% of genotypes are new, with just residual overlap to the previous sample (Fig.\ \ref{fig:Num-nets}c4). Abundance profiles readily move away from $g_{\rm l}$ (gray curves in Fig.\ \ref{fig:sim_distances}a). Despite the apparent fragmentation, relocation is a replacement of the root, not a loss of structure. The network remains self-similarly hierarchical at all times. 

A picture of two timescales emerges in all three cases: (i) rebuilding the hierarchical network around a root is fast and always happens; (ii) root relocation is slow, stochastic, and increasingly rare as adaptation progresses. 

\section{Empirical hepatitis B virus genotype networks}
\label{sec:hbv}

\begin{figure}[t]
    \centering
    \includegraphics[width=\linewidth]{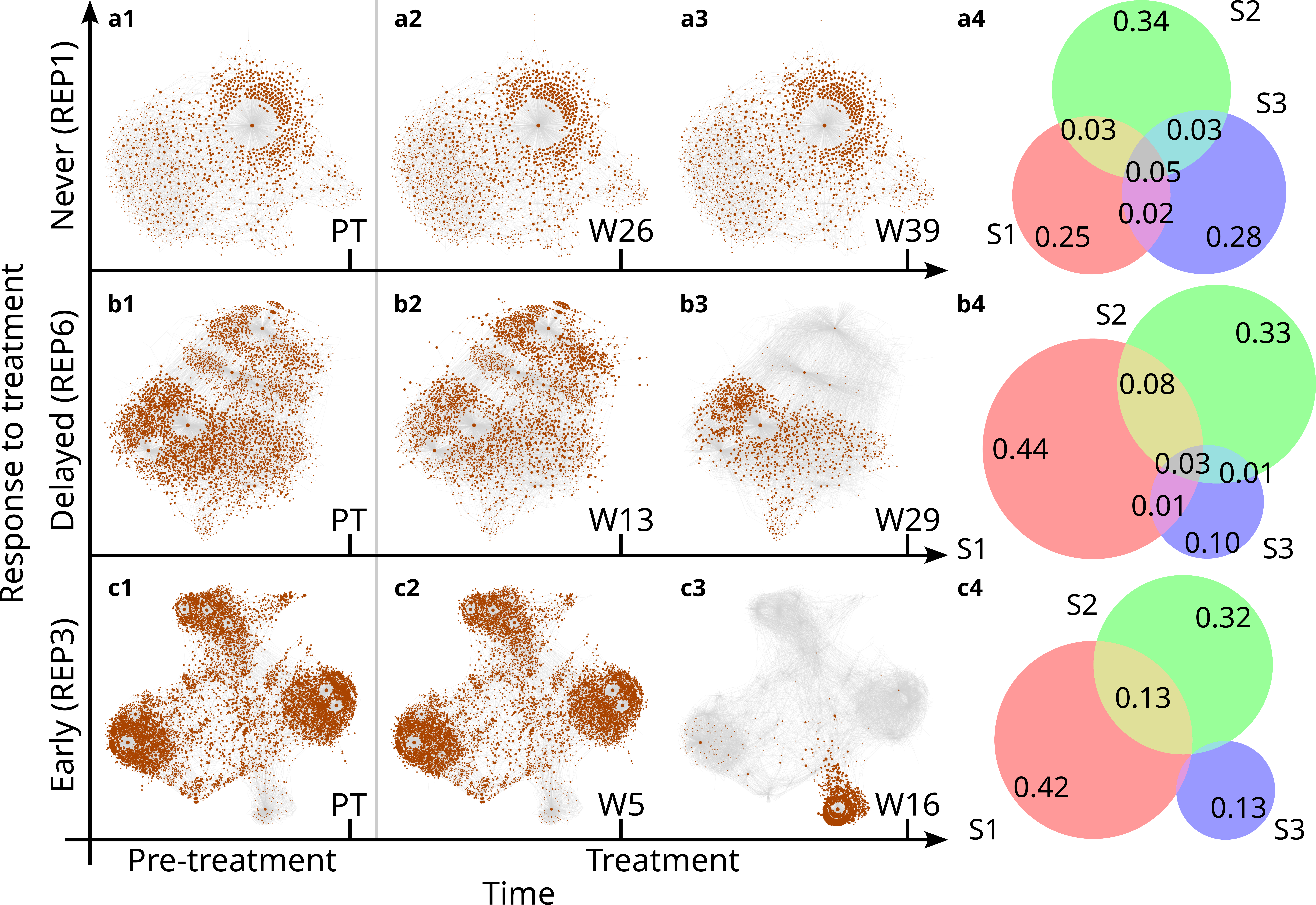}
    
    \caption{{\bf Genotype networks from empirical data.} We show three representative cases---in order of how promptly they respond to treatment: ({\bf a}) never, REP1; ({\bf b}) over an intermediate time, REP6; or ({\bf c}) early, REP3. This order defines an axis for these three patients, which are echoed by apparent network structure revealed through the panels. But this should be taken with care, and does not generalize well to all patients. Empirical cases present more subtleties than the model can accommodate---which are discussed in the text. Each patient was sampled at three time points (S1, S2, and S3) which differed between individuals. Nodes with non-zero abundance are shown in orange at S1 ({\bf a1}, {\bf b1}, and {\bf c1}), S2 ({\bf a2}, {\bf b2}, and {\bf c2}), and S3 ({\bf a3}, {\bf b3}, and {\bf c3}). Node size is proportional to the logarithm of the abundance. All edges (even to inactive nodes) show the network layout at all times. Venn diagrams ({\bf a4}, {\bf b4}, and {\bf c4}) show the fraction of all sequences that are present at each individual sampling, and at intersections between samplings (note that the intersection between two samplings must include the fraction present in all three).} 
    
    \label{fig:HBV-nets}
\end{figure}

\begin{figure}[t]
    \centering
    
    \includegraphics[width=0.8\textwidth]{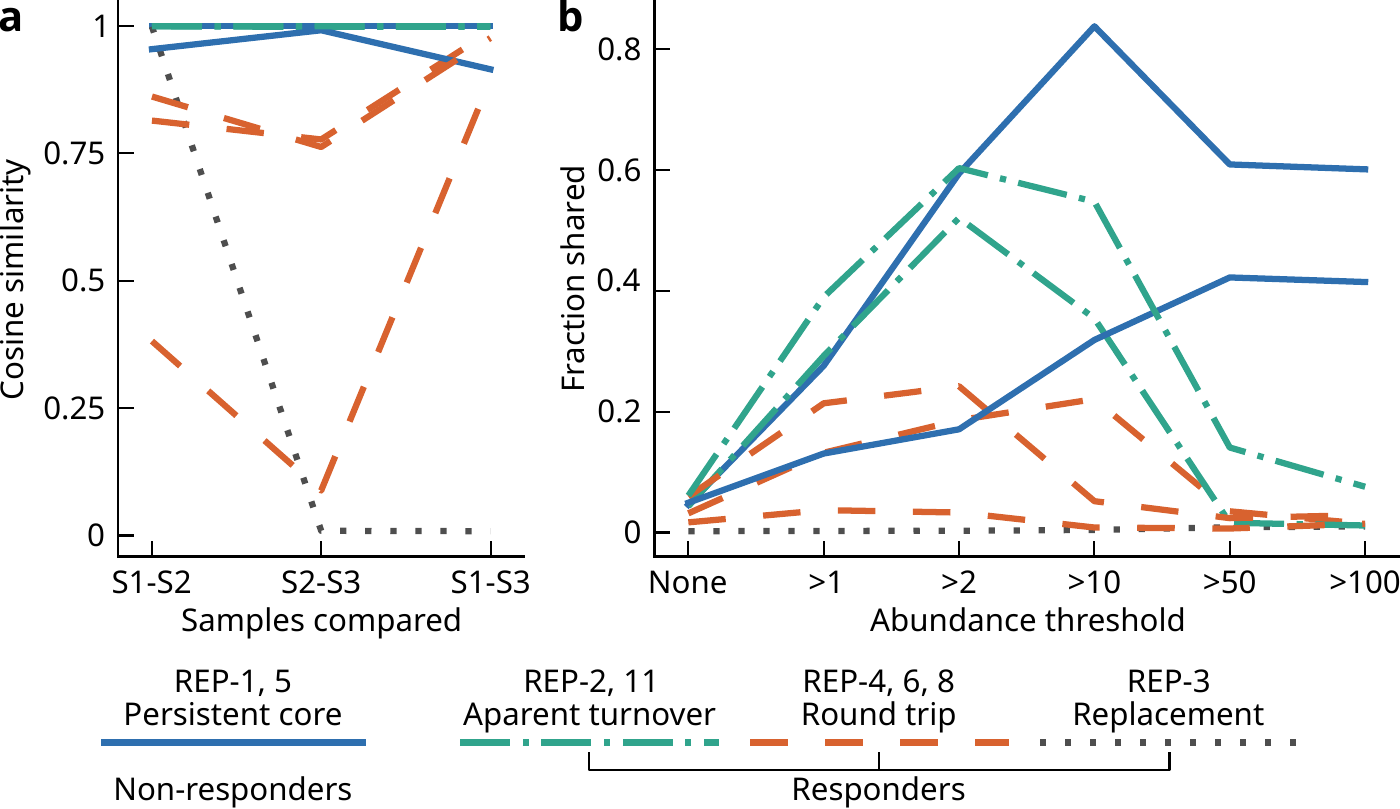}
    \caption{{\bf Quantifying overlap between longitudinal samples of each patient.} ({\bf a}) Cosine similarity is dominated by persistence or replacement of the root, due to the heavy tailed nature of abundance distributions in empirical viral quasispecies. ({\bf b}) To get a view of turnover not dominated by the root, we quantify the fraction of shared sequences across all samples as we threshold them by abundance. Both indicators are needed to get a comprehensive picture of turn-over in heavy-tailed distributions, and together they reveal different dynamical classes across genotype networks in our patients.}
    
    \label{fig:response-groups}
    
\end{figure}

The eight patients analysed (Table~\ref{tab:patient_summary}) underwent therapy with REP~2139-Ca but had different outcomes. Six showed marked reductions of viral load: early and rapid in REP3, REP8 and REP11, progressive in REP2 and over an intermediate period in REP6. Patients REP1 and REP5 were non-responders. Some organization principles cut across conditions: all networks are hierarchical and have broad, heavy-tailed degree distributions at every sampling (Fig.~\ref{fig:sim_distances}c), as in previously analyzed empirical populations~\cite{seoane:2026,martinezalcala:2026}. Other features differ and, since treatment is essentially the same, differences in network organisation can be put in correspondence to different responses. Fig.\ \ref{fig:HBV-nets} shows the quasispecies evolution for patients REP1, REP6, and REP3, with time to response to treatment roughly defining an ordering (vertical axis), which we can compare to simulation outcomes. This is only approximate: patients REP2 and REP11 do not fit nicely into this picture. Combining evidence from cosine similarity and overlap plots (Fig.\ \ref{fig:response-groups}), we suggest that HBV genotype networks fall within three dynamical kinds, and that two of them correspond to the stasis (Fig.\ \ref{fig:Num-nets}a) and relocation (Fig.\ \ref{fig:Num-nets}c) regimes from simulations, while a third case falls in-between and is not clearly captured by the model. 

\subsection{Stasis of the root: REP1}

REP1 and REP5, both non-responders, are the empirical counterparts to the global-root simulations. REP1 network is organised around the same root at S1, S2 and S3 (Fig.~\ref{fig:HBV-nets}a1-3), and the abundance profiles of its three samples fall on top of each other, decaying by about five orders of magnitude over six mutations (blue curves in Fig.~\ref{fig:sim_distances}b), as in the global run (Fig.~\ref{fig:sim_distances}a). All three samples share around $5\%$ of genotypes (Fig.~\ref{fig:HBV-nets}a4) and the cosine similarity between samples is essentially one (as for REP5, solid blue in Fig.~\ref{fig:response-groups}a). Most shared sequences in both patients have high abundance (solid blue in Fig.\ \ref{fig:response-groups}b) indicating that stasis of the root coexists with turnover of a low-abundance periphery---again, a sign of replication-exploration balance near a peak. 

The network of responders REP2 and REP11 also retains the root. Cosine similarity stays high (above $0.9$, dot-dashed turquoise in Fig.\ \ref{fig:response-groups}a), which is only explained by a constant root. But overlap as a function of sequence abundance (dot-dashed turquoise in Fig.\ \ref{fig:response-groups}b) suggests a complex story. It is low when considering all sequences, compatible with periphery turnover. It raises as we dismiss less abundant nodes, suggesting a non-negligible persistent core. But it finally collapses for the most abundant sequences, indicating that, except for the root, most hubs are indeed replaced. Since REP2 and REP11 are both responders, this strongly suggests that minority variants can influence the phenotype of a viral quasispecies \cite{borderia:2015}, impacting on response to treatment beyond preservation of the root (often conflated with a consensus sequence). Changes in the accompanying variants may contribute to a response to therapy even when the dominant clone remains unchanged. 

\subsection{Excursion and return of the root: REP6}

REP6, a responder, shows a behaviour that our simulations, by construction, cannot capture. This is revealed by the abundance profile (Fig.\ \ref{fig:sim_distances}b), which starts centered around the root for S1 (dotted orange curve), then broadens and has its maximum displaced at S2 (dashed orange), and finally returns and stays closer to the root at S3 (solid orange). Cosine similarity is accordingly reduced whenever S2 is involved, but remains high otherwise (dashed orange in Fig.\ \ref{fig:response-groups}a). This back and forth is also visible over the network (Fig.\ \ref{fig:HBV-nets}b1-3). Overlap as a function of abundance (dashed orange in Fig.\ \ref{fig:response-groups}b) resembles that of REP2 and REP11, suggesting a similar complex story that again highlights the impact of quasispecies structure on treatment response beyond preservation of the root sequence. REP4 and REP8 (with the strongest transient divergence) follow the same pattern. 

The present simulations cannot test what produces this pattern, but the mismatch between model and data suggests a hypothesis. Simulated sequences have $L=250$ sites, all of them observed. In turn, the amplicon spans $339$ of an about $3200$ nucleotides genome. It is a projection onto a small fraction of the dimensions relevant for adaptation. A transient excursion of the full-genome quasispecies, driven by mutations that become fixed elsewhere in the genome, would appear in the amplicon as a temporary departure from, and return to, the original root. These additional mutations can change, through epistasis, the fitness value of sequences in the observed genome fragment. This reading is consistent with the data, but so is a treatment-driven change of the fitness landscape. Without linkage information with the rest of the genome, our data cannot discriminate between them. 

\subsection{Relocation of the root: REP3}

REP3, a rapid responder, presents features from both the evolved and local-root simulations. Between S1 and S2 the quasispecies is stable: the two abundance profiles coincide (discontinuous gray curves in Fig.~\ref{fig:sim_distances}b; but note already a visible mode far from S1's root, see below) and their cosine similarity is $\sim1$ (dotted gray in Fig.\ \ref{fig:response-groups}a). By S3 the population has been almost entirely replaced: the cosine similarities involving S3 drop to near-orthogonality (dotted gray in Fig.~\ref{fig:response-groups}a) and the fraction of nodes shared by all samples is below 1\% at every abundance threshold (dotted gray in Fig.~\ref{fig:response-groups}b). The abundance profile at S3 is a new hierarchy whose maximum lies four mutations away from the S1 root, while the abundance at the old root has dropped by about two orders of magnitude (Fig.~\ref{fig:sim_distances}b). Both metrics indicate a fast replacement of the dominant genotype rather than gradual drift, which is the signature of simulations with evolved, $g_{\rm e}$, and local, $g_{\rm l}$, starting conditions: 

a stasis of certain time length followed by the rebuilding of the cloud around a root a few mutations away (solid gray in Fig.~\ref{fig:sim_distances}b). 
The second mode mentioned above in S1 and S2 abundance profiles (discontinuous curves in Fig.~\ref{fig:sim_distances}b), located around $6$ or $7$ mutations from the root, might foreshadow a shift to a new root. Several hubs are indeed visible on the network (Fig.~\ref{fig:HBV-nets}c1-2)---a state resembling the extended search in genotype space starting from local optima, $g_{\rm l}$ (Fig. \ref{fig:Num-nets}c).

\section{Discussion and conclusions}
\label{sec:discussion}

Our results support a picture of quasispecies adaptation as a transit between metastable states that unfolds on two timescales. On the faster timescale, the population rebuilds its diversity and hierarchical network around the current root. This hierarchy is present at every sampling time in all simulated and empirical populations (Figs.~\ref{fig:Num-nets} and~\ref{fig:HBV-nets}). On the slower timescale, the population stochastically finds a new local maximum and relocates around it. In simulations, the abundance profile remains fixed when the root is the global maximum, shifts by a few mutations after a long period of stasis when the root is an evolved genotype, and continues moving away from the initial genotype when the root is a poor local maximum (Fig.~\ref{fig:sim_distances}a). This picture reframes punctuated dynamics \cite{eldredge:1972} at a resolution finer than consensus-sequence turnover and helps explain how local search remains efficient in astronomically large spaces \cite{manrubia:2026}: populations do not diffuse through the entire space (potentially aided by neutral networks), but repeatedly re-anchor around a root, with each reorganisation occurring rapidly and locally.

Root turnover in our simulations was faster when the initial root was less well adapted (local $>$ evolved $>$ global, with no relocation from the global maximum). This ordering is qualitatively consistent with diminishing-returns, in which fitter backgrounds offer fewer and smaller adaptive opportunities, causing adaptation to decelerate \cite{chou:2011,khan:2011,kryazhimskiy:2014,wiser:2013}. In NK landscapes, this pattern arises naturally from ruggedness: the fitter a genotype, the fewer fitter genotypes are likely to lie within reach \cite{kauffman:1987,draghi:2013}. Global epistasis also provides a general framework for diminishing returns in complex traits \cite{reddy:2021}. We regard this connection as suggestive rather than conclusive. The ordering reflects the simulations shown, and relocation time might depend not only on fitness difference between the old and new roots, but on local landscape structure and stochastic search. 
Therefore, it may vary across landscape realisations. Moreover, because our roots are local maxima, improvement requires crossing a valley through the periphery of the population cloud \cite{weissman:2009}, rather than fixing a single beneficial substitution. Testing this analogy will require measuring relocation times and fitness gains as functions of root fitness across many landscape realisations. 

The HBV data show patterns analogous to the simulated regimes. The two non-responders (REP1 and REP5) and two responders (REP2 and REP11), in which low- and intermediate-abundance variants turned over, retained essentially the same dominant clone, as in simulations starting from a high-fitness root. In REP2 and REP11, however, the S3 populations differed in composition from those at S1 despite having the same root. This difference is not captured by the global-root simulations, in which no hidden variables affect the fitness of the observed genotypes. One responder (REP3) was almost entirely replaced by an unrelated dominant clone, consistent with relocation in the evolved- or local-root simulations (Fig.~\ref{fig:sim_distances}b). Three responders (REP4, REP6 and REP8) showed a transient shift of the dominant clone at the intermediate sample, followed by a return to the original root. For REP6, the abundance profile for S3 was similar to S1 (Fig.~\ref{fig:sim_distances}b).

These observations also highlight a difference between the simulations and the empirical data. The simulated sequences have $L=250$ sites, all of which are observed, whereas the HBV amplicon spans $339$ nucleotides of a genome of about $3\,200$, thus capturing only a small fraction of the dimensions relevant to adaptation. The roots measured in the HBV data are consequently roots of this projection. Mutations outside the amplicon could alter, through epistasis, the fitness landscape represented by the observed fragment. Such hidden changes could explain why REP2 and REP11 show different variant compositions despite retaining the same root. They could also produce an excursion of the full-genome population that appears in the amplicon as a temporary departure from, and return to, the original root. Conversely, a shift in the full-genome root may be invisible in the amplicon. The excursion-and-return pattern has no counterpart in our simulations, which makes this discrepancy informative. Alternatively, the pattern could reflect a treatment-driven change in the fitness landscape—for example, immune or drug escape that favours a genetic background only after an interacting mutation arises. The available data cannot distinguish between these explanations.

Root dynamics varied qualitatively among patients receiving the same treatment, but clinical response did not map one-to-one onto root movement. Two responders showed no root movement within the sampled period; this may reflect the collective action of the variant ensemble on adaptation to the new environment, an effect shown for Coxsackie virus B3 \cite{borderia:2015}. We therefore interpret root behaviour as associated with, but not predictive of, treatment outcome. Further studies are needed to determine how quasispecies organisation and the composition of its variants affect treatment response before a clear association can be established.

In summary, hierarchical local search combined with rare relocation may be a general strategy by which molecular populations meet challenges such as temperature change, immune pressure or antiviral treatment without prior knowledge of the fitness landscape. This architecture has been reported in RNA-virus populations studied in vitro and in naturally infected populations \cite{seoane:2026,martinezalcala:2026}, and now in a DNA virus \emph{in vivo}, supporting the possibility that it is shared across these systems. Whether root dynamics and genotype-network topology can serve as clinical readouts remains to be determined. A natural next step is longitudinal deep sequencing of larger genome fragments, with denser and more homogeneous sampling in a larger cohort.

\section*{Acknowledgments}

The authors acknowledge support from the Spanish State Research Agency, AEI, through grants PID2023-147963NB-C21 (IAD, SMA, SM) and PID2023-153225NA-I00 (LFS) funded by MICIU/AEI/10.13039/501100011033 and ERDF, EU. LFS received funding from the Occident Foundation (grant FJSCNB-2022-12-B) and worked on this manuscript while visiting the Okinawa Institute of Science and Technology (OIST) through the Theoretical Sciences Visiting Program (TSVP). 

\bibliographystyle{apsrev4-2}
\bibliography{bibliography}

\end{document}